# Photoinduced iodide repulsion and halides-demixing in layered perovskites

Yongtao Liu[1,2], Miaosheng Wang[2], Anton V. Ievlev[1], Mahshid Ahmadi[2], Bin Hu[2], Olga S. Ovchinnikova[3]*

1. Center for Nanophase Materials Sciences, Oak Ridge National Laboratory, Oak Ridge, Tennessee 37830, United States
2. Joint Institute for Advanced Materials, Department of Materials Science and Engineering, University of Tennessee, Knoxville, Tennessee 37996, United States
3. Computational Science and Engineering Division, Oak Ridge National Laboratory, Oak Ridge, Tennessee 37830, United States

* Author to whom correspondence should be addressed.

Olga S. Ovchinnikova

Computational Science and Engineering Division

Oak Ridge National Laboratory

1 Bethel Valley Rd

Oak Ridge TN, 37831-6493

ovchinnikovo@ornl.gov.

**Abstract**

Mixing halides in metal halide perovskites (MHPs) is an effective approach to adjust MHPs bandgap for applications in tandem solar cells. However, mixed-halide (MH-) MHPs undergo light-induced-phase-segregation (LIPS) under continuous illumination. Therefore, understanding the mechanism of LIPS is necessary for developing stable MH-MHPs. In this work, we investigated LIPS in layered (L) MHPs and discovered a critical role of spacer cations in LIPS. Through probing chemical changes of LIPS, we unveil light-induced-iodide-repulsion and the formation of Br-rich-phase in illuminated regions during LIPS. This discovery also gives insight into LIPS process in three dimensional (3D) MHPs. By further investigating LIPS in 3D MHPs, we reveal that LIPS induces not only the formation of Br-rich and I-rich domains but also an overall change of halide distribution along the film thickness direction, which can affect the electronic energy alignment and consequently MHPs devices performance. Moreover, LIPS is more significant in the bulk due to larger population of photogenerated charge carriers. Overall, this study reveals the chemical mechanism of LIPS in MHPs and its potential effect on device performance, offering insight into understanding LIPS mechanism and improving the stability of MHPs.

Metal halide perovskites (MHPs) community has witnessed a rapid growth in the development of photovoltaics in the past decade, with a power conversion efficiency (PCE) increase from 3.8 % to 25.5 % in single-junction MHPs solar cells.[1] MHPs-based multijunction solar cells are now attracting enormous research interests to further boost the PCE of MHPs solar cells.[2-5] The possibility of tuning bandgap via adjusting and mixing halides[6] makes MHPs a promising candidate for applications in tandem solar cells with various structures, such as MHPs/MHPs,[7] organic photovoltaics/MHPs,[8] silicon/MHPs,[2, 4] and Copper Indium Gallium Selenide (CIGS)/MHPs multijunctions.[9] However, mixed-halide MHPs undergo phase segregation under photoillumination,[10-12] which is detrimental to achieving stable photovoltaic devices. Therefore, understanding the process of light-induced phase separation (LIPS) will be a vital step for application of mixed-halide MHPs in stable solar cells.

LIPS in MHPs was discovered by Hoke et al.,[12] showing as a bandgap transition and photoluminescence (PL) redshift of mixed-halide MHPs under continuous illumination. Later, extensive researches investigated how LIPS affects optical properties of MHPs, such as emission, absorption, and so on.[10, 11, 13-16] LIPS originates from the formation of separated iodine ($I^-$) rich and bromine ($Br^-$) rich domains.[10, 11, 13-17] Therefore, LIPS is a chemical change due to halide migration driven by illumination. However, to date, direct observation of chemical changes due to LIPS in MHPs is missing. This is because LIPS is a reversible process,[10, 11, 13-16] which recovers (remixing of separated I-domain and Br-domain) quickly in several minutes or hours when the samples store in dark condition, leading to challenges in probing the chemical changes directly. For this reason, irreversible LIPS processes will be good model systems to investigate chemical changes associated with LIPS and hence understand the underlying mechanism.

Currently, to improve the stability of MHPs, large spacer cations have been widely used, which lead to the formation of layered (L) MHPs structure and play roles in suppressing ion migration.[18, 19] In addition to improve the stability of traditional 3D MHPs, L-MHPs have also established themselves as an excellent material for the application of light-emitting field. In this regard, the emission color of L-MHP can also be controlled via mixing halides, resulting in mixed-halides L-MHP and consequently raising a question regarding the LIPS in mixed-halides L-MHP. However, to date, most research into LIPS mainly focused on 3D-MHPs,[20-27] understanding of LIPS in layered MHPs is currently missing.

In this work, we investigated LIPS in layered MHPs (L-MHPs). We reveal that the spacer cation in L-MHPs plays a crucial role in LIPS, suggesting that adjusting cation is an effective approach to manage LIPS. In contrast to reversible LIPS in 3D-MHPs that recovers quickly under dark, LIPS in L-MHPs is irreversible and lasts for several days. This allows us to directly probe the chemical redistribution due to LIPS, which has not been thoroughly studied yet. Here, we use time-of-flight secondary ion mass spectrometry (ToF-SIMS) to probe the chemical distribution in MHPs. We find that illumination repels iodide ions and induces the formation of Br-rich in the illuminated region. Iodide repulsion is dependent on the light intensity, which results in a variation of halide distribution in both vertical and lateral directions of MHPs films. We further reveal that the chemical environment (e.g., oxygen penetration into MHPs) also affects LIPS. To comprehensively explore chemical changes due to LIPS in an operando condition, we illuminated a MAPb(Br$_{0.5}$I$_{0.5}$)$_3$ (MA: methylammonium) film by a uniform white light LED and probed its chemical changes in both vertical and lateral directions. In addition to the formation of I-rich and Br-rich domains that has been discussed extensively, we also found a significant vertical halide redistribution due to LIPS, which can play critical roles in charge carrier transport, recombination, and extraction in these materials. Moreover, we discovered that LIPS occurs more significantly in the bulk of MHPs film than the surface, which is probably due to a higher density of photogenerated charge carriers in the bulk.

**Results and Discussions**

We investigated LIPS in L-MHPs with different spacer cations, as shown in Figure 1a-c. Although previous works investigated the effect of cations (e.g., MA, formamidinium (FA), cesium (Cs)) on LIPS, the choice of cation is very limited in 3D-MHPs and these cations are still very similar in size and structure (e.g., MA *vs.* FA). In contrast, the choice of cation in L-MHPs is broader, offering a platform to better understand the role of cation in LIPS. We selected three spacer cations with different sizes and different functional groups (Figure 1a-c), namely, ethylammonium (EA), n-butylammonium (BA), and phenethylammonium (PEA). Using these cations we synthesized L-MHPs with an equal amount of bromide (Br) and iodide (I). The studied materials are, $EA_2Pb(Br_{0.5}I_{0.5})_4$, $BA_2Pb(Br_{0.5}I_{0.5})_4$, and $PEA_2Pb(Br_{0.5}I_{0.5})_4$, abbreviated as EA-, BA-, and PEA-MHP, respectively. Photoluminescence (PL) spectroscopy was first used to investigate LIPS in these L-MHPs. PL is a widely utilized spectroscopy to investigate LIPS phenomena in MHP.[12, 16] LIPS shows as a continuous red-shift of PL peak due to the formation of separated Br-rich and I-rich phases; photogenerated carriers in Br-rich phase transfer to I-rich phase before recombining and emitting (as shown in Figure S1). As a result, PL spectra predominantly (or only) show the emission peak of the I-rich phase (that is red-shifted compared to the original emission peak of mixed Br-I phase) indicating chemical changes after LIPS.

In our PL measurement, we used a 375 nm laser with an intensity of 54.7 mW cm$^{-2}$ to illuminate and excite the L-MHP samples. We observed a significant red-shift in the PL peak wavelength of EA-MHP under continuous illumination, as shown in Figure 1d-e. This PL behavior of EA-MHP (Figure 1d-e) is very similar to that of 3D-MHPs confirming LIPS in this sample. The original PL intensity gradually decreases and a newly emerged PL peak (in longer wavelength) corresponding to I-rich phases gradually increases. However, we only observed a small red-shift of ~1 nm in the PL peak position in BA-MHP sample (Figure 1f-g). In addition, we observed an emergence of a small peak at ~407 nm (insert of Figure 1g) that corresponds to the PL emission of pure-Br phase (i.e., $BA_2PbBr_4$) in BA-MHP. Accordingly, there is also LIPS in BA-MHP sample, which is evidenced by the formation pure-Br phase (insert of Figure 1g). However, the LIPS in BA-MHP is very weak as the overall changes in PL spectrum of BA-MHP are very small. In drastic contrast to EA-MHP and BA-MHP, we did not observe any obvious PL change in PEA-MHP (Figure 1h-i), suggesting the PEA-MHP is stable to illumination, and does not show any signs of LIPS in our measurement condition.

We further explored the effect of illumination intensity on the LIPS in these three L-MHPs. The PL spectra of three L-MHPs with different excitation intensities are shown in Figure S2-S4. We used a lower intensity (12.5 mW cm$^{-2}$) and a higher intensity (155 mW cm$^{-2}$) 375 nm laser to illuminate L-MHP samples and studied their PL behavior as a function of time. All observed phenomena in the three L-MHPs samples are similar to those we observed in Figure 1: (1) EA-MHP shows a red-shift of PL peak, (2) BA-MHP shows a rise of new PL peak around 407 nm corresponding to pure-Br-phase BA-MHP, and (3) PEA-MHP shows no obvious change in the PL spectrum after 30 mins continuous illumination. In the EA-MHP sample, the red-shift in PL is weaker under lower illumination intensity (Figure S2a-b) and stronger under higher illumination intensity (Figure S2c-d). This suggests that LIPS in EA-MHP is illumination-intensity-dependent. In the BA-MHP sample, the emergence of a PL peak around 407 nm indicating a Br-rich-phase[28] is also illumination-intensity-dependent (shown in the inserts of Figure S3b and S3d), suggesting the LIPS in BA-MHP is also illumination-intensity-dependent. The illumination-intensity-dependency of LIPS in EA-MHP and BA-MHP samples is consistent with previous observations in 3D-MHPs,[27] indicating the similarity of LIPS between L-MHPs and 3D-MHPs. Therefore, the underlying mechanism of LIPS in 3D-MHPs and L-MHPs can be similar, suggesting that understanding LIPS in L-MHPs will also offer useful knowledge for LIPS in 3D-MHPs.

Noteworthily, no obvious change in PEA-MHP sample is observed even with highest illumination intensity here (Figure S4c-d), suggesting PEA-MHP shows a higher resistance to LIPS than both EA-MHP and BA-MHP, indicating the role of spacer cation in the stability of L-MHP.

Interestingly, under higher illumination intensity (155 mW cm$^{-2}$), a peak around 450 nm rises in EA-MHP PL spectrum (Figure S2d), which is associated with the pure-Br EA-MHP (i.e., $EA_2PbBr_4$) and is consistent with the behavior of BA-MHP. To confirm the formation of pure-Br EA-MHP, we further increased the illumination intensity to 570 mW cm$^{-2}$ (375 nm laser) to study the evolution of EA-MHP PL spectrum. As shown in Figure S5, we observed an obvious PL peak around 450 nm after 30 min illumination, verifying the formation of pure-Br-phase during LIPS in EA-MHP.

A distinct difference between LIPS in L-MHPs (EA- and BA-MHPs) and 3D-MHPs is its reversibility: LIPS in 3D-MHPs is reversible while LIPS in L-MHPs is irreversible. As shown in Figure S6, LIPS induced PL changes are still very stable after keeping the L-MHPs samples under dark condition for 1 hour. It is well known that LIPS is a chemical phenomenon, which is associated with photoinduced ion redistribution. However, owing to the reversibility of LIPS in

3D-MHPs, it is challenging to directly detect the chemical changes due to LIPS. Thus, most previous studies only focused on investigating the changes in optical properties due to LIPS. In contrast, irreversible LIPS in L-MHPs offers us an opportunity to directly detecting chemical changes, allowing us to understand the exact chemical changes and the detailed mechanism of LIPS. Irreversible LIPS in L-MHPs is likely due to the higher activation energy for ion migration in L-MHPs, therefore, automatic ion redistribution under dark condition is inhabited in L-MHPs.

Next, we study the nature of chemical changes in L-MHPs due to LIPS using ToF-SIMS. We used a rectangular-shape laser spot (375 nm) to illuminate the L-MHPs samples, where the rectangular width is less than the maximum field of view of ToF-SIMS (500 x 500 μm). Therefore, we can cover both the illuminated and non-illuminated regions in a ToF-SIMS dataset, this allowed us to directly compare the chemical changes associated with illumination. The rectangular-shape illumination is shown in Figure 2a, where the blue stripe represents the illuminated region. Since we only observed LIPS in EA-MHP and BA-MHP, we only discuss chemical changes in EA-MHP and BA-MHP here. After laser illumination (30 min), we transferred the samples to ToF-SIMS for chemical characterization.

To explore the effect of illumination intensity on chemical changes, we varied the laser intensity and performed chemical characterization after various intensity illumination. According to Beer-Lambert law, we expect the chemical change in the vertical direction to be depth-dependent due to the decay of laser intensity inside the materials. Note that the laser used here is not perfectly uniform, the intensity of this laser decays from the middle to the edge of the rectangular due to gaussian profile of the laser intensity within the laser spot. Therefore, depending on whether or not the laser can penetrate the MHP films, the expected illumination conditions and corresponding resultant chemical distributions under strong illumination (can penetrate) and weak illumination (cannot penetrate) are schematically shown in Figure 2b and 2c, respectively. Previous works revealed that laser illumination can expel iodide in MHPs into solvent when the MHP sample was in contact with a solvent.[29] Our PL results show that strong laser illumination can lead to the formation of pure Br-phase (the appearance of PL from pure Br-phase, Figure 1 and Figure S5), also implies an iodide-repulsion effect. Therefore, we speculate that the laser treatment results in an I-deficient region in the illumination region. Based on this analysis, the expected chemical distributions after strong and weak illuminations are also schematically shown in Figure 2b-c.

ToF-SIMS results of Br and I distributions in BA-MHP and EA-MHP samples are shown in Figure 2d-g. The top row of each panel shows the integrated top-view (x-y) chemical distribution maps and the bottom-row of each panel shows the integrated cross-section-view (x-z) chemical distribution maps; the coordinate (x-y or x-z) is labeled at the bottom-right of each row. In Figure 2d, we observed an I-deficient (top-right of Figure 2d) and Br-rich (top-left of Figure 2d) stripe in the illuminated regions, suggesting that illumination expels iodide and results in I-deficient and Br-rich (IDBR) regions. In the cross-section-view (bottom row of Figure 2d), we find that I-deficient and Br-rich region is slightly broader at the top (near surface), which is consistent with our expectation in Figure 2b, suggesting laser penetration indeed plays a role in LIPS chemical changes. The laser of stripe-edge is too weak to penetrate the film and hence only affect chemical changes near the film surface. When we reduce the laser intensity, the laser of the stripe-middle is not able to penetrate to the film and only affects the chemical changes near the surface. This case is shown in Figure 2e, from the cross-section-view of Br and I distribution (bottom row of Figure 2e), we can see that the IDBR effect near the bottom middle is very weak. In addition, the overall chemical distribution in this case (bottom row of Figure 2e) is perfectly consistent with our schematic analysis in Figure 2c. Also, the IDBR effect is apparently weaker under the weaker illumination (Figure 2e) than that under stronger illumination (Figure 2d), suggesting that the iodide-repulsion effect is light-intensity-dependent.

The IDBR stripe in the illuminated region is also observed in EA-MHP, indicating that illumination similarly repels iodide in the EA-MHP sample. However, the cross-section view (bottom row of Figure 2f-g) reveals significantly different distributions of the chemicals. One can see broadening of IDBR region close to the bottom interface with substrate. This IDBR distribution is inconsistent with predicted ion distribution (Figure 2b-c), which implicates unconsidered factors affecting the iodide-repulsion during illumination.

Interestingly, in the cross-section-view oxygen distribution maps in Figure 2h-i (oxygen from glass substrate can penetrate to the thin MHPs), we observed a relatively uniform oxygen distribution in BA-MHP (Figure 2h), without any correlation with the IDBR region. In contrast, the oxygen distribution in EA-MHP (Figure 2e) shows correspondence to the IDBR region and is dependent on light intensity. This correlation between oxygen distribution and IDBR region (as marked in Figure 2e) indicates that oxygen in MHP also affects iodide-repulsion. It has been previously revealed that oxygen in MHPs can lead to an up-shift of conduction band but has a negligible effect on valence band.[30] Accordingly, an oxygen-rich condition (which is near the substrate and in illuminated region, as shown in Figure 2i) in EA-MHP will lead to an electronic band bending and results in hole-rich area in the corresponding region, as schematically shown in Figure 2j. It is revealed that the hole carriers have a role in repulsion iodides in MHPs.[29, 31] Therefore, we propose that the correlation between oxygen distribution and IDBR region is due to oxygen-induced hole accumulation, which repels iodide. In summary, under illumination, oxygen from substrate penetrates to the EA-MHP and leads to a local hole-rich environment which results in iodide-repulsion (we refer to this phenomenon as oxygen-induced iodide-repulsion). This oxygen-induced iodide-repulsion, together with light-induced iodide-repulsion, results in the abnormal IDBR regions in EA-MHP (Figure 2f-g).

The difference in oxygen distribution in BA-MHP (Figure 2h) and EA-MHP (Figure 2i) also indicates that spacer cations (e.g., BA vs. EA) effect on oxygen penetration in MHPs, which is important to MHP performance and stability. Therefore, the role of spacer cation in ion migration and stability of MHPs is worth further exploring in the future.

It has been suggested that the formation of Br-rich and I-rich domains (as schematically shown in Figure 3a) due to LIPS detrimentally affects the photovoltaic performance of MHPs by trapping charge carriers in I-domains.[32] Our chemical investigations show that LIPS can also induce a chemical change along vertical direction of MHP films (as schematically shown in Figure 3a), which will significantly affect charge carrier transport and extraction. In LIPS studies above (Figure 1-2), we illuminated MHP samples by a laser, which is not a uniform illumination. Thus, an IDBR region corresponding to the illuminated region formed due to light-induced iodide-repulsion. However, in an operando condition, MHP films should be under a uniform illumination (that is, the whole MHP film is under illumination). Therefore, we next systematically investigate light-induced chemical distribution change in MHP under a uniform illumination in order to understand light-induced chemical changes in operando devices. We utilized a built-in white LED light panel in ToF-SIMS chamber to illuminate MHP samples, as the LED illumination area is much larger than MHP sample size, this illumination can be treated as a uniform illumination.

We performed ToF-SIMS measurements on a MAPb($Br_{0.5}I_{0.5}$)$_3$ sample under both dark and LED illumination to understand light-induced chemical changes. ToF-SIMS results are shown in Figure 3b-d. In the ToF-SIMS depth profiles (Figure 3b), we clearly observed a difference in Br and I distribution between dark (solid curves) and illumination conditions (dashed curves), indicating light-induced halide redistribution along the vertical direction. We also simultaneously investigated halide distribution in lateral direction, shown as chemical maps in Figure 3c-d. Under illumination, we clearly observed microscale Br-rich and I-rich domains (Figure 3d) in comparison with dark condition (Figure 3c), suggesting light-induced formation of Br-rich and I-rich phases. Therefore, LIPS results in not only the formation of Br-rich and I-rich domains but also overall changes of halide distribution along vertical direction. Noteworthily, effective charge carrier transport in a MHP device is along the vertical direction, therefore, LIPS induced vertical halide redistribution will affect charge carrier transport and MHP devices' performance. This is consistent with recent literature;[33] which revealed that an improper energy alignment due to LIPS is are responsible for the photovoltaic performance loss.[33] Here, our results on vertical redistribution of halides elucidates the chemical origin of this improper energy alignment. The energy alignment related to halide redistribution will affect charge carrier transport and consequently photovoltaic performance of MHPs.[33]

In the depth profiles in Figure 3b, we explicitly see that light-induced changes in halide distribution mainly located at the bulk of MHP film, suggesting that MHP's bulk undergoes most significant LIPS. Therefore, we further analyzed the lateral halide segregation in the bulk and surface of MHP separately. As shown in Figure S7a, we show the lateral distribution of Br and I near surface and bulk of the MHP film. Consistent with our speculation, a more significant light-induced halide segregation (exhibits as the formation of Br-rich and I-rich domains) is seen in bulk (Figure S7c) than near surface (Figure S7b). Since it has been revealed that halide migration and segregation are associated with charge carriers,[29, 31] the more significant LIPS in bulk is probably because of a larger population of photogenerated charge carriers at bulk than the surface.

**Conclusions**

In conclusion, we investigated LIPS in L-MHPs, unveiling that spacer cation plays a crucial effect on LIPS. We observed irreversible LIPS in L-MHPs; this irreversible LIPS offers an opportunity to directly investigate chemical redistribution due to LIPS, which is challenging in 3D MHPs that show quickly reversible LIPS behavior. Our chemical characterization of LIPS in L-MHPs revealed an iodide-repulsion effect and consequently the formation of pure-Br-phase or extremely Br-rich phase in the illuminated region, where iodide repulsion is dependent on light intensity and chemical environment (e.g., oxygen penetration into MHPs affect iodide repulsion process). In order to understand chemical redistribution due to LIPS in an operando condition, we illuminated a MAPb(Br$_{0.5}$I$_{0.5}$)$_3$ film using a uniform white LED light. In addition to the formation of I-rich and Br-rich domains (that is a generally believed mechanism of LIPS), we discovered that LIPS also leads to a significant vertical chemical redistribution, which is critical to charge carrier transport and extraction in MHPs devices. Moreover, we discovered that LIPS is more significant in the bulk of the MHP film, which is probably due to a large population of photogenerated charge carriers in the bulk. Overall, by directly characterizing chemical changes due to LIPS, we for the first time revealed the chemical changes associated with LIPS in MHPs, which is the key for understanding LIPS mechanism and developing stable mixed-halides MHPs. Our studies about the role of spacer cation in LIPS and extrinsic ion penetration also offers insight into improving the stability of MHPs.

## Acknowledgments

This research was supported by Center for Nanophase Materials Sciences at Oak Ridge National Laboratory. This research was partially conducted at the Center for Nanophase Materials Sciences, which is a DOE Office of Science User Facility and using instrumentation within ORNL's Materials Characterization Core provided by UT-Battelle, LLC under Contract No. DE-AC05-00OR22725 with the U.S. Department of Energy.

## Author contributions

Y.L. conceived the project and O.S.O directed the experiments. Y.L. prepared the samples. Y.L. performed PL measurements with help from M.W.. A.V.I. performed ToF-SIMS measurement. All authors contributed to discussions of results and manuscript.

## Competing Interests

The authors declare no competing financial interests.

## Methods

**Materials preparation**: The metal halide perovskites samples were synthesized on Swiss glass slides. The glass slides were pre-cleaned using deionized water, acetone, and isopropanol. $EA_2Pb(Br_{0.5}I_{0.5})_4$ were synthesize by spin-casting a precursor (1 M $PbI_2$ and 2 M EABr in d dimethyl sulfoxide) at 3000 rpm, then, the spin-casted film was annealed at 100 ºC for 10 min. $BA_2Pb(Br_{0.5}I_{0.5})_4$ were synthesize by spin-casting a precursor (1 M $PbI_2$ and 2 M BABr in d dimethyl sulfoxide) at 3000 rpm, then, the spin-casted film was annealed at 100 ºC for 10 min. $PEA_2Pb(Br_{0.5}I_{0.5})_4$ were synthesize by spin-casting a precursor (1 M $PbBr_2$ and 2 M PEAI in d dimethyl sulfoxide) at 3000 rpm, then, the spin-casted film was annealed at 100 °C for 10 min. MAPb(Br$_{1.5}$I$_{1.5}$) was synthesized by spin-casting a precursor (0.8 M lead (II) acetate trihydrate, 1.2 M MAI, and 1.2 M MABr in dimethylformamide) at 3000 rpm, then, the spin-casted film was annealed at 100 °C for 30 min. Precursor preparation, film spin-casting, and annealing were performed in a N$_2$-filled glove box.

**Photoluminescence (PL) measurement:** PL measurements were performed by using Horiba Fluorolog-3 spectrometer. The excitation source is from a 375 nm continuous-wave laser combined with the focus lens. Filters were used to adjust laser intensity and laser shape.

**Time-of-flight secondary ion mass spectrometry (ToF-SIMS):** ToF-SIMS measurements were performed using ToF-SIMS.5.NSC instrument (ION.TOF GmbH. Germany). Experiments were carried out in negative ion detection mode using Bi$_3^+$ primary and Cs$^+$ sputter ion sources. Bi$_3^+$ primary ions source with energy of 30 keV, current of 0.5 nA and spot size of ~120 nm was used to generate secondary ions of the studied sample. Sputter source with energy of 500 eV, current of 70 nA and spot size ~20 μm was used for depth profiling and removing of the significant amount of material for chemical studies in the bulk. Measurements were carried out in non-interlaced mode, where each chemical scan (500 x 500 μm, 256 x 256 px, ~20 s duration) with primary source was followed by sputtering (750 x 750 μm, 4 s duration). Time-of-flight mass analyzer was used to analyze negative secondary ions with mass resolution m/Δm = 100 – 500.

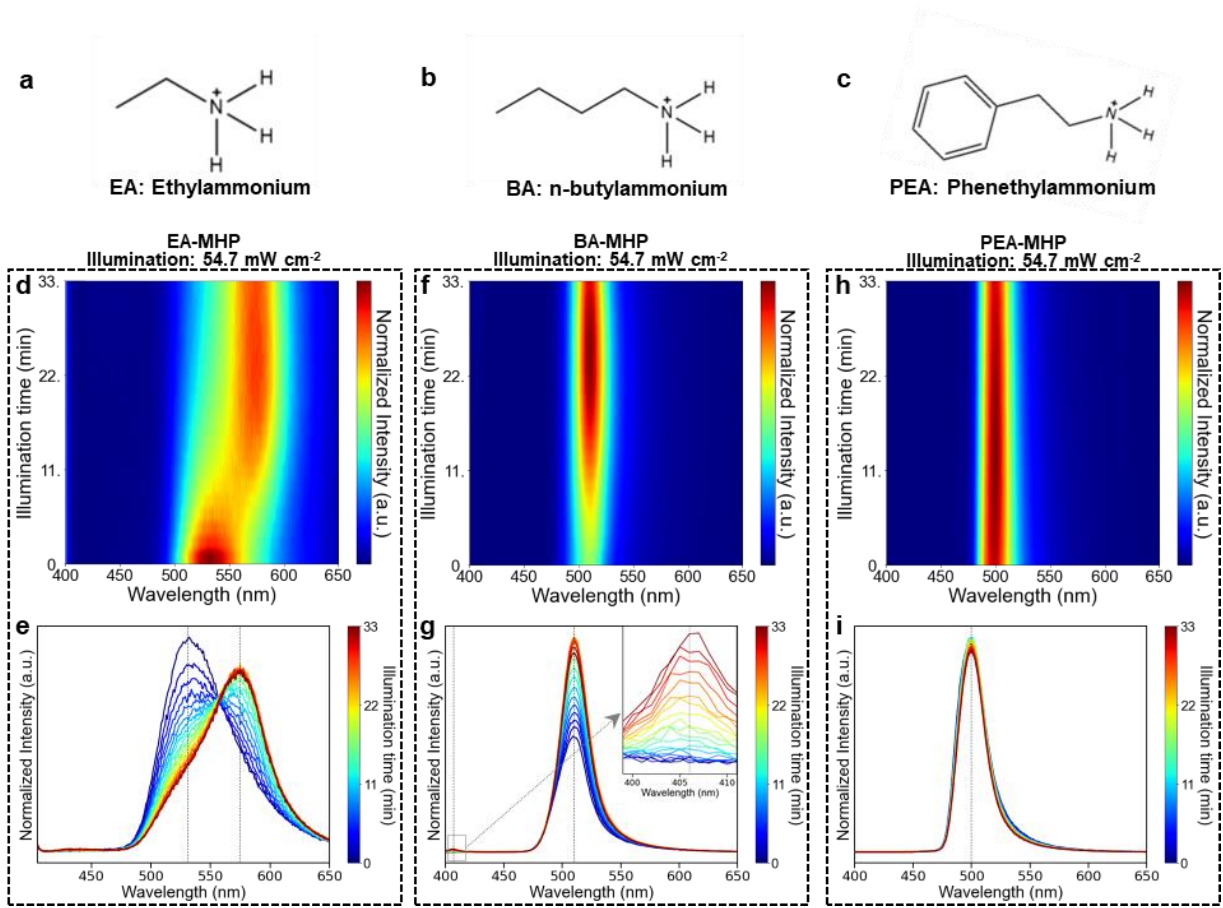

**Figure 1. a-c,** molecular structures of three spacer cations, **a,** EA; **b,** BA; **c,** PEA. **d, f, h,** contour images of PL evolution under continious illumination of three L-MHP; **d,** EA-MHP; **f,** BA-MHP; **h,** PEA-MHP. **e, g, i,** PL spectra corresponding to contour images in **d, f, h,** respectively.

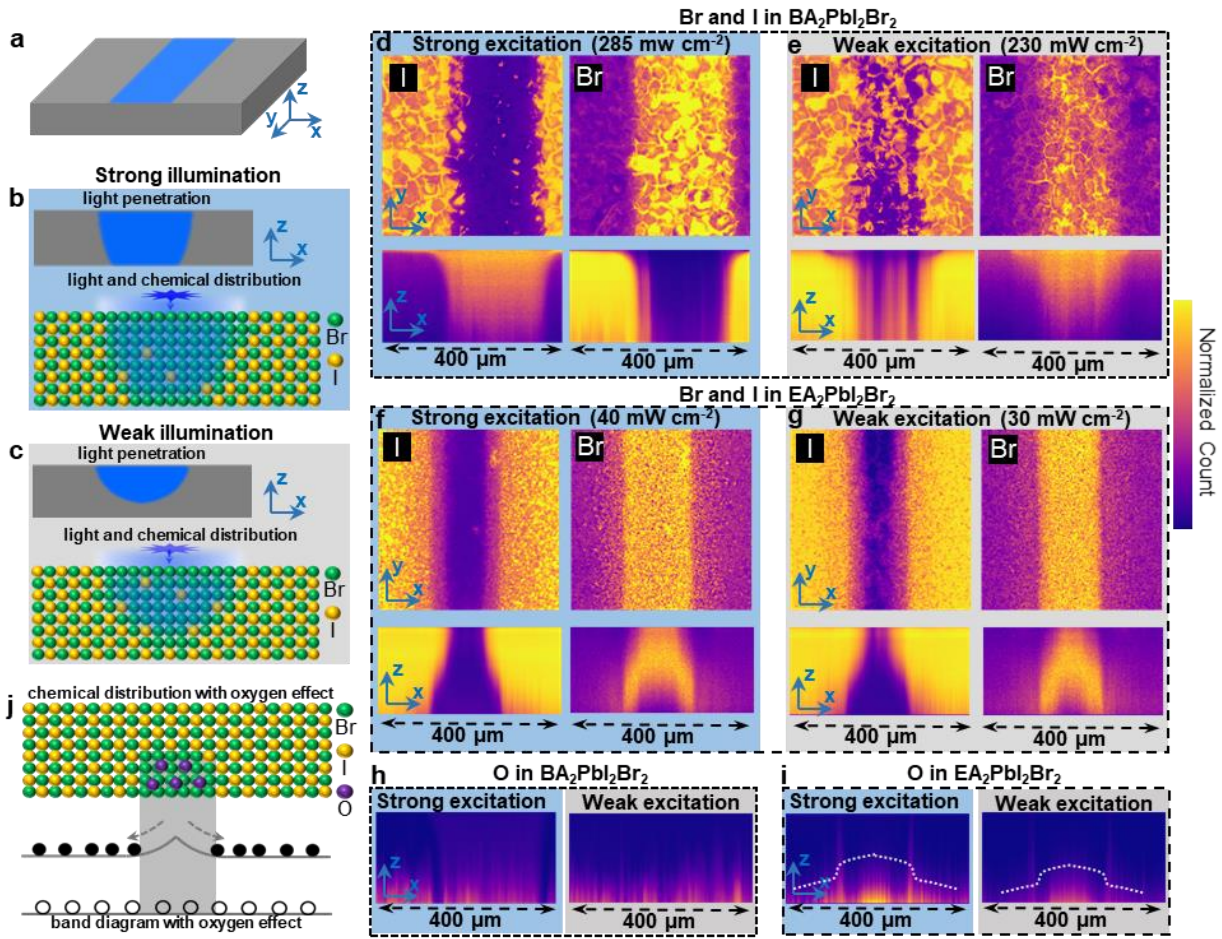

**Figure 2. a,** schematic of stripe-shape illumiantion, the blue stripe represents illuminated region. **b,** cross-section-view of expected illumination condition under a strong illumination (top) and corresponding halide distribution. **c,** cross-section-view of expected illumination condition under a weak illumination (top) and corresponding halide distribution. **d,** I and Br distribtuions in BA-MHP under strong illuminaiton. **e,** I and Br distribtuions in BA-MHP under weak illuminaiton. **f,** I and Br distribtuions in EA-MHP under strong illuminaiton. **g,** I and Br distribtuions in EA-MHP under weak illuminaiton. In **d-g**, top-row is integrated top-view and bottom-row is integrated cross-section-view. **h,** oxygen distribution in BA-MHP after strong and weak illumination. **i,** oxygen distribution in EA-MHP after strong and weak illumination

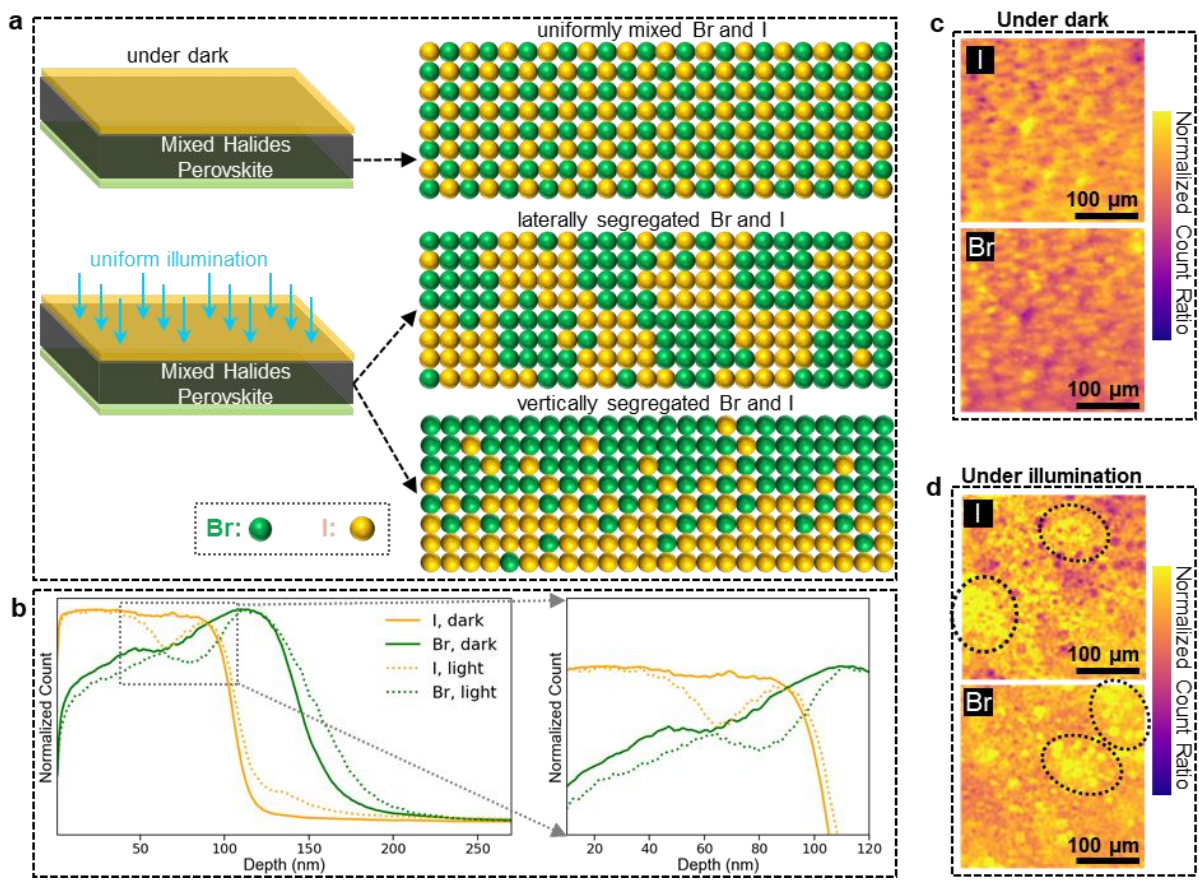

**Figure 3. a,** schematic of expected illumination condition and halide distribution under dark and operando condition. **b,** depth profile of iodide, bromide, and oxygen distribution in MAPb(Br$_{0.5}$I$_{0.5}$)$_3$ under dark (solid lines) and illumination (dashed lines), a significant change in Br and I distribution due to illumination is observed. **c,** lateral distribution of I and Br under dark was shown as the ratio; the ratio was calculated as: e.g. iodide ratio, $R_I$ = (I count)/(I count + Br count). **d,** lateral distribution of I and Br under illumination. Microscale I and Br segregation is observed under illumination (**d**) compared to those under dark (**c**), the I-rich and Br-rich regions are marked by circles in I ratio map and Br ratio map in (**d**), respectively.